# DPI-SPR: A Differentiable Physical Inversion for Shadow Profile Reconstruction Framework in Forward Scatter Radar

ShuQi Lei[1], *Member, IEEE*, Gan Yu[2], Yuan Tian[3] and XiaoWei Shao1*
Shanghai Jiao Tong University, 800 Dongchuan Rd, Shanghai, 200240, China
sqlei99@sjtu.edu.cn, ty-tina@sjtu.edu.cn, dx_zhang@sjtu.edu.cn; *shaoxwmail@163.com;

*Abstract*—Forward scatter radar (FSR) has emerged as an effective imaging modality for target detection, utilizing forward scattering (FS) signals to reconstruct two-dimensional shadow profile images of objects. However, real-world FS signals are inevitably corrupted by noise. Due to the ill-posed nature of electromagnetic inversion and its high sensitivity to noise, existing imaging methods often suffer from degraded performance or even complete failure under low signal-to-noise ratio (SNR) conditions. To address this challenge, we propose DPI-SPR (Differentiable Physical Inversion for Shadow Profile Reconstruction), an end-to-end imaging paradigm built upon the Secondary Wave-Source Response Field (SWRF). The core concept of this paradigm is to reformulate the imaging problem as an optimization problem of continuous and learnable SWRF parameters. To this end, we develop a fully Differentiable Forward Scattering model (DFSM). Leveraging this model, the proposed inversion framework integrates a robust logarithmic loss with physics-based regularization constraints, enabling accurate gradient propagation from observation errors to the SWRF parameters associated with the shadow profile. Extensive simulation experiments have been conducted to verify the effectiveness and robustness of the inversion proposed framework. The results show that our method achieves high-precision profile reconstruction directly from limited and noisy reference signals, even at an SNR as low as 8dB, setting a new benchmark for robust imaging in FSR scenarios.

## I. INTRODUCTION

Forward scatter radar is a specialized bistatic radar system that operates with a near-180° observation geometry[1]. This configuration greatly enhances diffraction, resulting in an order-of-magnitude increase in radar cross-section (RCS)[2]. Consequently, the FSR is ideal for detecting low RCS or stealth targets[3-6]. It has also been used to image target shadow profiles.

However, the lack of range resolution in FSR geometry limits its imaging capabilities to only two-dimensional target shadow profiles projected onto a plane perpendicular to the transmitter-receiver baseline[7, 8]. Reconstructing accurate shadow profiles based on one-dimensional noisy signals is a complex electromagnetic inverse scattering problem[9]. As shown in Fig. 1, the effectiveness of existing methods is usually affected in the case of noisy or distorted FS signals[10-12]. Therefore, in order to promote the practical applications of FSR, it is crucial to develop a highly robust target shadow profile reconstruction method.

Currently, the mainstream research directions for FSR shadow profile imaging mainly include two types of technical paths: shadow inverse synthetic aperture radar (SISAR) imaging[13] and profile retrieval techniques based on the forward scatter shadow ratio (FSSR)[14].

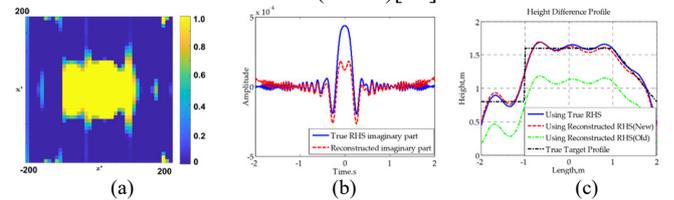

**Fig. 1** Shadow profile image of existing methods. (a) Shadow profile imaging of rectangular target used samples with 1/4 of the total pixels, under SNR=10dB[10]. (b) The red line is the imaginary part of distorted RHS. (c) The green line is the height difference profile with size distortion[12].

### A. SISAR imaging for FSR

In 2000, with the first proposal of SISAR[13], FSR imaging achieved a key breakthrough. SISAR uses the motion of the target along the baseline to synthesize a large effective aperture. The basic theory of SISAR is to combine doppler compensation with the inverse Fresnel transform of the radio holography signal(RHS) to recover the complex profile function(CPF) of target[13]. The final profile is derived from the magnitude and phase information of the CPF. However, the distortion of the RHS will cause the reconstructed shadow profile with scale distortion as shown in Fig. 1(b)(c). Therefore, the segmented Hilbert transform and additional low-pass filtering are introduced to improve the accuracy and consistency of RHS reconstruction[16].

In order to adapt to different diffraction angles, related studies have made targeted corrections to the SISAR signal model and Doppler parameter estimation algorithm[15-18]. Under the conditions of long baseline and small diffraction angle, the theoretical resolution of SISAR is about 2.5λ to 5λ [9]. With the development of dual-frequency large diffraction angle imaging technology(nearly $180° \pm 50°$), the resolution has been improved to 0.8λ~2.4λ[17]. Furthermore, based on GNSS opportunity signals and using a modified SISAR signal model, the experimental results found that scale distortion existing in the shadow profile of aircraft targets by normal Hilbert transformation[11].

A series of imaging algorithms based on SISAR theory are essentially analytical modeling, and their effectiveness is highly dependent on three key assumptions: accurate diffraction angle prediction and specific signal model, constant target velocity, and precise Doppler parameter estimation and RHS reconstruction. In practical applications, if these assumptions are not met, the performance can easily be severely degraded. For example, for near-field targets, Doppler information may be severely lost, resulting in a significant decrease in imaging capabilities[11].

In summary, breaking through the strong dependence of existing SISAR imaging models on diffraction angles and scenes, and improving the ability to robustly and accurately reconstruct profiles under RHS signal distortion or noise conditions are core scientific problems that need to be solved in the field of FSR imaging.

*B. Using FSSR for shadow profile retrieval*

A more recent approach in passive FSR imaging is based on the FSSR, which connects the ratio of received to incident power to the shadow profile of the target theoretically[14]. The paper[19] first realized the shadow profile retrieval of the target using a set of rectangular strips based on hard thresholding and the least squares method. Subsequently, the method was further refined to the pixel level, enabling the retrieval of a more accurate target shadow profile with an effective resolution of 10 $\lambda$ on the grid side[20].

Early optimization methods required the number of FSSR samples to exceed the total number of pixels and did not account for noise effects[19-22]. Recent studies have proposed a denoising method that uses approximate gradient information rather than accurate gradient descent. While this approach enhances noise resistance in shadow profile retrieval and reduces the required number of samples[10], the reliance on approximate gradients can complicate the gradient propagation path and hinder stable convergence[23, 24]. As illustrated in Fig. 1(a), the shadow profile of a rectangular target retrieved by 400 FSSR samples with a SNR of 10dB in a 40 × 40 grid is shown.

More fundamentally, FSSR-based methods are derived from the approximate Fresnel diffraction model[25], limiting their physical accuracy in non-paraxial or near-field scenario. The optimization is also performed at a phenomenological level, where the learned pixel densities serve as a mathematical substitute for the target shape rather than representing a direct physical quantity. Critically, these methods leave unresolved the intrinsic non-differentiability of the forward scattering model, owing to the implicit step-function representation of the target boundary[10][19-22]. This limitation fundamentally precludes the application of modern optimization techniques based on automatic differentiation. Therefore, the need for proposing a novel shadow profile reconstruction method with high accuracy that is physically meaning and provides a robust and stable optimization process.

*C. Differentiable forward electromagnetic calculation*

In recent years, differentiable physical modeling has emerged as a new paradigm for solving electromagnetic inverse problems. Foundational work in optics first demonstrated that approximate gradients could guide optimization even for discontinuous physical processes[26]. This technical concept was rapidly extended to electromagnetic computation, leading to end-to-end optimization frameworks integrated with time-domain electromagnetic simulation[27] and the development of the FDTD equivalent convolutional neural network [28].

This approach also proved highly effective in radar imaging. Subsequent works introduced a differentiable SAR renderer that created a direct gradient path from scattering mechanisms to the image[29]. Recently, a differentiable ray tracing framework based on a differentiable rendering model has been implemented for end-to-end inversion from SAR images to surface scattering parameters[30].

These studies demonstrate that building a physically faithful and fully differentiable forward model is both necessary and efficient for tackling complex electromagnetic inverse problems with modern optimization methods. Inspired by these advances, we extend the differentiable physics paradigm to FSR.

*D. Our Contributions*

This paper proposes a DPI-SPR framework for reconstructing shadow profiles by inverting target characteristic parameters. By introducing learnable parameters with clear physical significance, a fully DFSM is developed. Recasting the classically ill-posed inverse problem as a continuous optimization task enables direct gradient propagation from measurement residuals to a physically interpretable representation of target geometry. Compared with existing FSSR or SISAR algorithms, this method shows significant improvements in generality, robustness, and imaging accuracy.

In summary, the key contributions of this work are as follows.

1) We introduce the secondary wave-source response field (SWRF), a continuous and learnable field that characterizes the shadow casting properties of the target. This allows us to reformulate the DFSM to depend on the SWRF, decoupling the integration domain from the unknown geometric structure through a probabilistic approximation, thereby resolving the inherent non-differentiability issue of classical diffraction models.

2) We develop a robust end-to-end DPI-SPR framework that reformulates the ill-posed FSR imaging problem as an unconstrained SWRF parameter optimization problem. A composite loss function is designed to integrate a logarithmic loss of FS signals with specialized binary entropy and geometric regularization priors, ensuring a stable optimization process that converges to physically plausible and structurally consistent profiles, even under severely corrupted FS signals.

3) We conduct extensive simulations to quantitatively validate the effectiveness of the proposed framework. Using the intersection over union (IoU)[31] and the 95th percentile Hausdorff Distance(95%HD) metrics[32], we achieve accurate reconstruction of both convex and non-convex shadow profiles. Our method also exhibits strong robustness, achieving high accuracy even under a SNR of 8dB in FS signals affected by mixed Gaussian and impulse noise. The majority boundary errors of the reconstructed profiles remain below 1.5$\lambda$, calculated using the 95%HD.

The paper aims to promote the development of FSR imaging technology towards the fusion of physical modeling and artificial intelligence. The rest of this paper is organized as follows: Section II discusses the details of the DPI-SPR framework. Section III presents qualitative and quantitative experimental results. Section IV concludes the article.

## II. PROPOSED METHOD

The proposed FSR imaging paradigm is implemented through a differentiable physical inversion framework (DPI-SWRF) as shown in Fig. 2. This framework consists of a forward differentiable physical modeling and a backward gradient-based parameter learning process. This section will provide a detailed introduction.

In Section II-A, we present a novel DFSM based on a learnable and continuous SWRF. This model is elaborated in Section II-A, addressing the non-differentiability issue inherent in classical diffraction theory, enabling the inverse problem to be solved using gradient optimization methods.

The SWRF parameters are subsequently learned through an end-to-end optimization process. As described in Section II-B, this process is guided by the gradients computed from a delicately designed composite loss function, which includes robust data fidelity terms and physical priors to ensure stable convergence.

Finally, section II-C describes the process of reconstructing the shadow profile from the optimized SWRF. This critical step transforms the continuous SWRF distributed across the aperture plane into a physically reasonable and deterministic binary profile through probability mapping and thresholding operations.

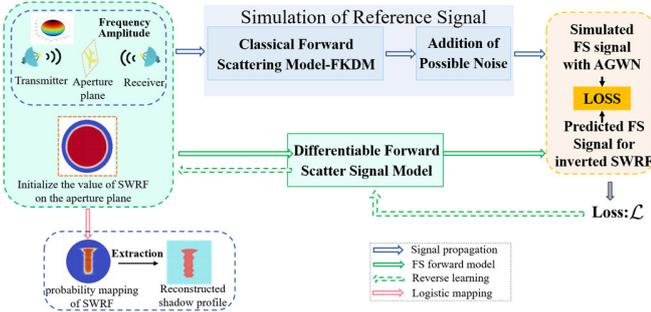

Fig. 2. Shadow Profile Reconstruction framework

### A. Forward Scatter Model

The FSR system topology is illustrated in Fig. 3(a). The transmitter T is positioned at the origin of the coordinate $xyz$, with the positive direction of the $y$ axis defined as the vector from the transmitter to receiver R. The baseline length of $\overline{TR}$ is $L$. The target center traverses the baseline, moving from point $P_0$ to $P_N$.

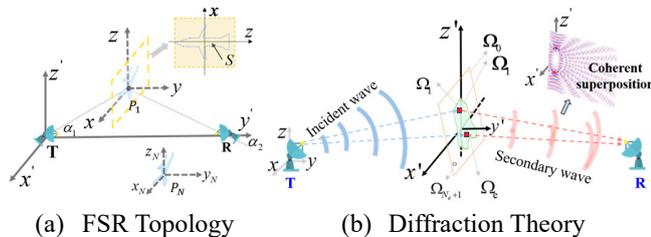

(a) FSR Topology  (b) Diffraction Theory
**Fig. 3.** Physical principles of the FS signal and FSR Topology

Where R captures and processes $N$ discrete FS signals. The coordinate $x'y'z'$ parallel to $xyz$, is established with $P_i$ as the origin. It is assumed that the target pose remains constant.

$\alpha_1$ and $\alpha_2$ represent the angles between the positive $y$-axis and the vectors $\overline{TP_i}$ and $\overline{P_iR}$.

Diffraction theory and the Babinet principle provide the physical foundation for modeling FS signal[13]. The Huygens-Fresnel principle provides a classical physical explanation for diffraction phenomena[33], which states that every point on the wavefront can be regarded as a new source of secondary spherical wave. The electromagnetic field at any point in space is the coherent superposition of these secondary waves.

Therefore, the complex amplitude at point R can be mathematically expressed as

$$\tilde{U}(R) = \iint_{wavefront} d\tilde{U}(R) \qquad (1)$$

Let $Q$ denote the center of an arbitrary secondary wavelet on the wavefront, and $d\Omega$ be the surrounding area. The contribution of each secondary spherical wave emitted from $Q$ to the field at R can be calculated as

$$d\tilde{U}(R) = KF(\alpha_1, \alpha_1)\tilde{U}_0(Q)\frac{e^{jkr_2}}{r_2}d\Omega \qquad (2)$$

The complex amplitude $\tilde{U}_0$ at the point $Q$ due to the primary source T is given by

$$\tilde{U}_0(Q) = \frac{e^{jkr_1}}{r_1} \qquad (3)$$

Here, $r_1, r_2$ denote the respective distances from transmitter T and the observation point R to each point $Q = (x', z')$.

Kirchhoff rigorously establishes that the integration can be limited to the diffraction screen $\Omega_1$ on the aperture plane $x'oz'$ rather than on the entire closed surface[34, 35]. This simplification greatly simplifies the mathematical representation of the diffraction problem. The Fresnel-Kirchhoff diffraction integral[36] provides a forward scattering field model with optimal theoretical accuracy, expressed as

$$\mathcal{F} = \frac{jA}{2\lambda}\iint_S \varepsilon_T(x',z')\frac{e^{jk(r_1+r_2)}}{r_1 r_2}(\cos\alpha_1 + \cos\alpha_2)dx'dz' \qquad (4)$$

Where the integration is over the closed shadow region $S$, which is the projection of the true geometry onto the $x'oz'$. $\varepsilon_T$ is the indicator function of $S$[37], defined as

$$\varepsilon_T(x',z') = \begin{cases} 1, (x',z') \in S \\ 0, (x',z') \notin S \end{cases} \qquad (5)$$

The diffraction constant $K$ and tilt factor $F(\alpha_1, \alpha_1)$ are derived strictly from wave optics theory by Kirchhoff.

A key limitation of FKDM lies in its integration domain, which is directly related to the target shadow profile $S$. This profile is defined by a step function on the aperture plane $x'oz'$, but this function is non-differentiable at points on boundary, making it impossible to calculate the gradient of continuous shape parameters. Therefore, modern gradient-based optimization methods cannot be used to solve the electromagnetic inverse problem in FSR.

### B. Differentiable Forward Scatter Model

To resolve the non-differentiability of the classical Kirchhoff model, we reformulate the forward scattering problem. The central idea is to decouple the diffraction integral from the hard, geometric boundary of the target. This is achieved by introducing the SWRF, denoted by $\rho(x',z')$ and defined on the aperture plane $x'oz'$, taking values freely in $\mathbb{R}$. The SWRF is a continuous, spatially distributed scalar field that quantifies the propensity of each point $Q=(x',z')$ on the aperture plane to respond to an incident wave and emit a secondary wave. A higher SWRF value signifies the stronger tendency of this point $Q=(x',z')$ to emit secondary wave.

To bridge the unconstrained and real-valued SWRF with the binary characteristic of a physical shadow profile, this paper employs a differentiable relaxation. This function maps the SWRF value at each point $Q=(x',z')$ to a smooth activation coefficient $\sigma(\rho)$, bounded between 0 and 1[38]:

$$\sigma(\rho) = \frac{1}{1+\exp^{-\rho}} \qquad (6)$$

The value of $\sigma(\rho)$ represents the activation probability of the secondary wave source at any given point $Q$. This mapping creates a smooth, differentiable soft profile in place of the step boundary of the classical model.

Physically, the SWRF links activation intensity to the target shadow profile. As the optimization converges, points with high SWRF values (leading to $\sigma(\rho) \approx 1$) represent strong activation intensity and lie inside the shadow boundary $S$, while points with low SWRF values ($\sigma(\rho) \approx 0$) represent weak activation intensity (leading to $\sigma(\rho) \approx 0$) and lie outside of it. This soft representation of the physical boundary based on the learnable parameters forms the foundation of our new DFSM.

The DFSM is constructed by reformulating the classical FKDM to be fully differentiable with respect to the shadow casting properties of target. The classical integral calculates the scattered field by integrating over a sharply defined non-differentiable aperture area $S$. Our key modification is to replace this hard binary integration domain with the continuous, differentiable SWRF parameters.

The proposed DFSM retains the fundamental components of the classical model, including the diffraction constant $K$ and tilt factor $F(\alpha_1, \alpha_1)$. The integral with the contribution of each point Q smoothly weighted by its activation probability. This ensures a differentiable transition at the physical boundary of the target shadow profile.

The resulting mathematical formulation for the DFSM is

$$\mathcal{F} = \frac{jA}{2\lambda}(\cos\alpha_1 + \cos\alpha_2)\iint_{\Omega_0} \sigma(\rho)\frac{e^{jkr_1}}{r_1}\frac{e^{jkr_2}}{r_2}d\Omega \qquad (7)$$

Where A denotes the incident wave amplitude, λ is the wavelength, and k=2π/λ is the wave number. The integration domain of Equation (8) is $\Omega_0$ as shown in Fig. 3(b), which completely encloses the real shadow boundary of $\Omega_1$. Assuming that the target center is $p(x_p, z_p)$ in $xoz$ coordinate of Fig. 3(b), we can calculate $r_1, r_2$ by

$$r_1 = \sqrt{(x'+x_p)^2 + y_p^2 + (z'+z_p)^2}$$
$$r_2 = \sqrt{(x'+x_p)^2 + (L-y_p)^2 + (z'+z_p)^2} \qquad (8)$$

The DFSM possesses two critical advantages that make it superior to classical approaches. First, it is fully differentiable with respect to the SWRF parameters. This property is essential as it reformulates the inversion problem of FSR into an unconstrained optimization problem, enabling the direct and efficient application of modern gradient-based algorithms. Second, the model is derived under general spherical wave conditions without relying on simplifying distance approximations. This physically robust formulation ensures the generalizability of DFSM. Consequently, it is suitable for diverse and complex signal modeling scenarios, including near-field, far-field, and conditions involving both large and small diffraction angles.

*C. SWRF Parameter Learning Based on Gradient Descent*

This section details the inverse learning process for the SWRF parameter vector defined on the computational domain $\Omega_0$. The inversion is formulated as a gradient-based optimization problem where the optimal SWRF is found by iteratively minimizing a composite loss function. The following subsections detail its construction, including data consistency term and the physical regularization prior, and the inverse learning process.

1) The Composite Loss Function and Data Consistency Term

The SWRF vector $\rho$ is learned by minimizing a composite loss function that is the core of our DPI-SPR framework. The composite loss function is given by

$$\mathcal{L} = L\left((\mathcal{F}(\rho), \tilde{\mathcal{F}})\right) + \lambda_{Ent}\psi_{Ent}(\rho) + \psi_{Geo}(\rho) \qquad (9)$$

Where the function $\mathcal{L}$ consists of three key components. The term L is a physical data consistency. The term $\psi_{Ent}$ is a binary entropy regularization and the term $\psi_{Geo}$ is a geometric regularization. The hyperparameter coefficient $\lambda_{Ent}$ is used to balance the contribution of regularization item. The most fundamental component is the data consistency term L, which quantifies the misfit between the complex FS signals $\mathcal{F}(\rho)$ predicted by DFSM and the observed reference signals $\tilde{\mathcal{F}}$.

The information of the FS signal varies significantly with noise and observation geometry. For observed signals with rich information, the optimization priority is to accurately fit the details to ensure finer target geometry. In contrast, for observed signals with rich information, the priority transforms to robust convergence by suppressing the influence of outliers.

To ensure the generalizability of the DPI-SPR framework under various conditions, we adopt the complex Huber loss. The function is adaptively controlled by a hyperparameter $\delta$ to adapt to different noise and residual[38, 39].

The modulus of the complex residual between predicted and observed FS signals is denoted by

$$|r| = |\mathcal{F}(\rho) - \tilde{\mathcal{F}}| \qquad (10)$$

For large residuals $|r|$, the function behaves like a robust L1 loss. For small residuals, the function transitions smoothly to a L2 loss. The data consistency term $\mathcal{H}_\delta$ acts on a set of the complex residual are expressed as

$$\mathcal{H}_\delta\left(\mathcal{F}(\rho), \tilde{\mathcal{F}}\right) = \begin{cases} \frac{1}{N}\sum_{i=1}^{i=N}\frac{1}{2}|r|^2, |r| \leq \delta \\ \frac{\delta}{N}\sum_{i=1}^{i=N}(|r| - \frac{1}{2}\delta), |r| > \delta \end{cases} \qquad (11)$$

The function $\mathcal{H}_\delta$ allows the framework to adaptively balance robustness and accuracy, enabling physically faithful reconstruction in different FSR scenarios.

The logarithmic transformation of the Huber loss $\mathcal{H}_s$ is a key design choice that synergistically improves convergence speed and noise robustness[39, 40], which is expressed as

$$L = \ln\{1 + [\mathcal{H}_s(\mathcal{F}(\boldsymbol{\rho}), \tilde{\mathcal{F}})]\} \quad (12)$$

This equation provides overall stability for any large residual by creating a gradient scaling mechanism. According to the chain rule, the gradient with respect to $\boldsymbol{\rho}$ is

$$\nabla_\rho L = \frac{1}{1 + [\mathcal{H}_s(\mathcal{F}, \tilde{\mathcal{F}})]} \nabla_\rho \mathcal{H}_s(\mathcal{F}, \tilde{\mathcal{F}}) \quad (13)$$

Where $1/(1+\mathcal{H}_s)$ as a key gradient scaling factor, it significantly suppresses large gradients to prevent unstable updates, and instead performs smoothly exploratory of gradient updating to stabilize the loss. For very small residual, the logarithmic transformation hardly changes the original gradient information $\nabla_\rho \mathcal{H}_s$. This advantage ensures that the framework can still achieve refined fine-tuning when converging to a physically reliable solution.

2) Regularization Priors for Robustness and Plausibility

For ill-posed electromagnetic inverse problems, data consistency term L alone cannot guarantee physically meaningful solutions. Our DPI-SPR inversion framework introduces two prior regularization terms. The binary entropy prior regularization $\psi_{Ent}$ ensures the stability of optimization algorithm, and geometrical regularization $\psi_{Geo}$ enforces the physically geometric plausibility of the converged parameters.

The DFSM is constructed with learnable SWRF parameters as variables, which is mapped to a smooth probability distribution $\sigma(\rho)$ via a sigmoid function to ensure the differentiability of the DFSM. Based on the chain rule, the gradient of data consistency term L includes the gradient of the $\sigma(\rho)$, which derivative is calculated by the following equation.

$$\nabla_\rho \sigma = \sigma \cdot (1 - \sigma) \quad (13)$$

However, true shadow profiles are physically sharp, requiring high SWRF values inside boundary and low values outside. According to Equation (13), the gradients of these points are all close to zero. Therefore, the design of $\sigma(\rho)$ introduces a potential gradient vanishing problem during the optimization process.

By introducing the binary entropy regularization to resolve the gradient vanishing problem effectively[41], which is given by

$$\psi_{Ent}(\boldsymbol{\rho}) = -\frac{1}{M_e} \sum_{e=1}^{M_e} \begin{bmatrix} \sigma(\rho_e) \cdot \log(\sigma(\rho_e)) \\ +(1-\sigma(\rho_e)) \cdot \log(1-\sigma(\rho_e)) \end{bmatrix} \quad (14)$$

Where $M_e$ is the total number of secondary wave source in computational region $\Omega_0$. Its gradient with respect to learnable SWRF parameters is calculated by

$$\nabla_\rho \psi_{Ent} = \frac{1}{M_e} \sum_{e=1}^{M_e} [\frac{\partial \psi_{Ent}}{\partial \sigma} \cdot \nabla_\rho \sigma] = \frac{1}{M_e} \sum_{e=1}^{M_e} [-\log(\frac{\sigma}{1-\sigma}) \cdot \nabla_\rho \sigma] \quad (15)$$

The total gradient consists of the data gradient $\nabla_\rho L$ and the weighted entropy gradient $\lambda_{Ent} \cdot \nabla_\rho \psi_{Ent}$. When $\sigma$ approaches 0 or 1, the data gradient disappears, while the term $\log(\sigma/(1-\sigma))$ diverges to infinity. This adversarial mechanism prevents the SWRF from saturating too early. The binary entropy term maintains active exploration during the optimization process, avoiding convergence to local optimal solution due to overly confident predictions in the early stages.

Although the entropy regularization $\psi_{Ent}$ ensures active optimization results, it cannot guarantee that the geometry mapped by the converged SWRF parameters is physically realistic. Inspired by computational imaging in wave propagation [42], we introduce a set of geometric regularization priors as $\psi_{Geo}$ in Equation (16), that encode a general understanding of real targets.

$$\psi_{Geo} = \lambda_s \psi_s(\boldsymbol{\rho}) + \lambda_{conn} \psi_{conn}(\boldsymbol{\rho}) + \lambda_{com} \psi_{com}(\boldsymbol{\rho}) \quad (16)$$

It includes smoothness, connectivity, and compactness, which work together at different scales to enforce physically reasonable reconstruction. The details of these constraints are explained as follows.

a) Smoothness regularization is to eliminate high-frequency artifacts at the shadow boundary and promote the generation of smooth edge.
b) Connectivity regularization enhances spatial coherence by maximizing the similarity of activation intensity between two adjacent points, preventing the decomposition from splitting into discontinuous components.
c) Compactness regularization restricts the activation area to gather near the center of the target shadow.

$$\psi_s(\boldsymbol{\rho}) = \sum_{i,j} \mathbb{I}_{adj}(i,j) |\rho_i - \rho_j| / \sum_{i,j} \mathbb{I}_{adj}(i,j) \quad (17)$$

$$\psi_{conn}(\boldsymbol{\rho}) = -\frac{1}{M_e^2} \sum_{i \neq j} \frac{1}{1+(d_{ij}/\delta_c)^2} \cdot \sigma(\rho_i)\sigma(\rho_j) \quad (18)$$

$$\psi_{com}(\boldsymbol{\rho}) = \frac{1}{M_e} \sum_{i,j} (\sigma(\rho_i) d_{ij}) \quad (19)$$

Where hyperparameters $\lambda_s$, $\lambda_{conn}$, and $\lambda_{com}$ can be adjusted according to actual needs. The term $\mathbb{I}_{adj}(i,j)$ is adjacency matrix. The distance from each point to the origin of the local coordinate $x'oz'$ is denoted as $d_{ij}$.

The purpose of geometric regularization $\psi_{Geo}$ is to avoid post processing for optimization results. The three priors ensure that the final inverted profile is not just a fit to the data, but a structurally and topologically plausible, plausible representation of the physical object by placing constraints on the vast solution space[43, 44].

3) Gradient-Based Optimization

The optimal SWRF parameters $\boldsymbol{\rho}$ on computational region $\Omega_0$ are learned by minimizing the composite loss function $\mathcal{L}(\boldsymbol{\rho})$ via backpropagation of gradients iteratively.

$$\boldsymbol{\rho}^* = \arg\min_{\boldsymbol{\rho}} [\mathcal{L}(\boldsymbol{\rho})] \quad (18)$$

One of the main benefits of our DPI-SPR framework is its ability to define an unconstrained optimization problem for SWRF parameters. The formulation, shown in Equation (7), is elegant and allows for the full differentiability of SWRF parameters. This means that our end-to-end differentiable DPI-SPR framework can accurately compute gradients using automatic differentiation (AD) technology through backpropagation. By applying the chain rule, the total gradient is calculated as a weighted sum of the gradients from the composite loss. The process for calculating the gradient vector is as follows:

$$\begin{bmatrix} \frac{d\mathcal{L}}{d\rho_1} \\ ... \\ ... \\ \frac{d\mathcal{L}}{d\rho_{Me}} \end{bmatrix} = \frac{\partial L}{\partial \mathcal{F}} \begin{bmatrix} \frac{\partial \mathcal{F}}{\partial \rho_1} \\ ... \\ ... \\ \frac{\partial \mathcal{F}}{\partial \rho_{Me}} \end{bmatrix} + \lambda_{Ent} \begin{bmatrix} \frac{\partial \psi_{Ent}}{\partial \rho_1} \\ ... \\ ... \\ \frac{\partial \psi_{Ent}}{\partial \rho_{Me}} \end{bmatrix} + \lambda_{geom} \begin{bmatrix} \frac{\partial \psi_{geom}}{\partial \rho_1} \\ ... \\ ... \\ \frac{\partial \psi_{geom}}{\partial \rho_{Me}} \end{bmatrix} \quad (19)$$

We can concisely write it in matrix form as:

$$\frac{d\mathcal{L}}{d\rho} = \frac{\partial L}{\partial \mathcal{F}} \frac{\partial \mathcal{F}}{\partial \rho} + \lambda_{Ent} \frac{\partial \psi_{Ent}}{\partial \rho} + \lambda_{geom} \frac{\partial \psi_{geom}}{\partial \rho} \quad (20)$$

The analytical gradient path eliminates the need for complex and inefficient gradient estimation techniques. As a result, the inverse learning of $\rho$ can leverage mature gradient optimizers developed in deep learning frameworks.

In the forward propagation process, modern optimizers use DFSM to dynamically construct a computation graph. This graph consists of nodes and edges that record operations and dependencies. Then, during the backward process, the graph is traversed in reverse. This allows for the automatic calculation of gradients for the SWRF parameter vector using chain rules. These gradients are then efficiently propagated from the loss value. The entire process is illustrated in Fig. 4.

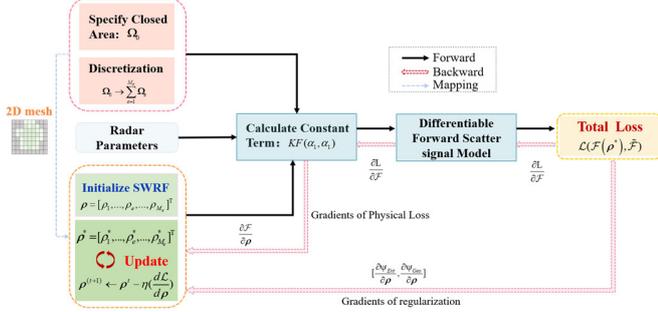

**Fig .4** Gradient propagation of continuous SWRF via AD

This process propagates the scalar values of the loss function backward, enabling efficient learning of the SWRF parameter vector. This end-to-end differentiable approach ensures the stability and robustness of the optimization process, enabling the framework to effectively learn the optimal SWRF distribution in $\Omega_0$ and reconstruct target shadow profile that are both physically reasonable and have high diffraction signal fidelity.

*D. Shadow Profile Reconstruction via Optimized SWRF*

This paper provides a new FSR imaging paradigm based on the modern AD optimizer. The preceding sections have detailed the theoretical components of our framework, including the DFSM and the composite loss function used for optimization of SWRF parameters. This section presents the practical implementation of the entire reconstruction process that consolidates the forward and backward propagation. The complete, step-by-step procedure, from the initial SWRF parameters to the final reconstructed shadow profile, is detailed in Algorithm 1.

**Algorithm1** Reconstruction Shadow Profile through Learning SWRF.

**Input:**
-Observed forward scattering complex field $\tilde{\mathcal{F}}$
-Initial SWRF parameter vector $\rho^{(0)} = [\rho_1, ......, \rho_{Me}]^{(0)}$
- learning rate $\eta$, regularization weights $\lambda_{Ent}, \lambda_{conn}$
Maximum number of iterations: $N_{iter}$

**Procedure:**
-for t=0 to $N_{iter} - 1$ do
　//---Forward Propagation---//
　- Substitute $\rho^{(t)}$ into DFSM to calculate $\mathcal{F}(\rho^{(t)})$
　- Substitute $\mathcal{F}(\rho^{(t)})$ into Eq. (11) to calculate L
　- Substitute $\rho^{(t)}$ into Eq. (14-17) to calculate $\psi$
　- Substitute $\lambda_{Ent}, \lambda_{conn}$ and L, $\psi$ into Eq. (9) to get $\mathcal{L}$
　//--- Backward Propagation---//
　- Compute the gradient of SWRF by Eq. (20): $\frac{d\mathcal{L}}{d\rho}$
　//---Updating---//
　-update SWRF using gradient descent: $\rho^{(t+1)} \leftarrow \rho^t - \eta(\frac{d\mathcal{L}}{d\rho})$
-end for
-Get the converged SWRF parameter vector: $\rho^* \leftarrow \rho^{(N_{iter})}$
//---Final profile Extraction---//
- compute activation probability map: $\sigma(\rho^*)$
- Applying a threshold in $\sigma(\rho^*)$ on $\Omega_0$:
$\Omega_{Tar} = \{(x',z') \in \Omega_0 \mid \sigma(\rho^*(x',z')) > 0.5\}$
-Reconstruct Kirchhoff boundary: $\Omega_{Tar} = 1, \Omega_0 - \Omega_{Tar} = 0$

**Output:**
Physically target shadow profile $\Omega_{Tar}$ in $\Omega_0$

The proposed framework is based on a loss function defined by SWRF parameters. By learning SWRF until the loss converges, we can obtain a relatively clear soft profile of target, which originates from probability mapping of activation intensities on secondary wave source. This allows us to accurately recover the final step-like hard boundary through a simple deterministic threshold operation.

In summary, the key contribution is the introduction of a fully differentiable, physics-driven forward model for the FS modeling and parameters inversion. By treating the DFSM as a differentiable solver, our framework achieves a soft-to-hard procedure. By first performing a stable optimization for a continuous physical SWRF and subsequently extracting the final binary profile, our framework provides a robust and physically faithful pathway to achieving high-accuracy FSR imaging.

## III. EXPERIMENTAL VALIDATION AND ANALYSIS

This chapter presents a comprehensive experimental validation of the proposed DPI-SPR framework. We introduce for the first time a rigorous quantitative evaluation scheme for FSR imaging. The reconstruction accuracy of the inverted profile is considered both globally and locally. The global reconstruction accuracy uses a pixel-level metric to evaluate the structural integrity, while the local geometric accuracy is evaluated using a boundary-based metric.

After a detailed description of the experimental setup in section A, we evaluate our approach along two key dimensions including generality and robustness. The generality of the DPI-SPR is tested on seven targets under two different observation scenarios. Subsequently, a robustness evaluation assesses the stability of the inversion framework when the reference signal is disturbed by severe noise with a SNR as low as 8dB. Finally, we conclude with a series of ablation studies to verify the specific contribution of each component to the final physically faithful reconstruction.

*A. Experimental Setup*

1) Simulation Parameters Setting

FSR applications include two typical observation scenarios: Far Field (FF) and Transition Zone (TZ). The FF scenario occurs when the target lies within the far-field regions of both transmitter and receiver. Conversely, the TZ scenario involves targets positioned in the near field of one antenna while remaining in the far field of the other. This configuration enables evaluation of near-field information exploitation for precise geometric inversion. The detailed simulation parameters for FS signals are summarized in Table 1.

TABLE I PARAMETERS OF SIMULATED REFERENCE FS SIGNALS

|  | Transition Zone | Far Field |
|---|---|---|
| Baseline Length $L$ | 500m | 10km |
| Crossing point $d_T$ | 100m | 3030m |
| Target width $D$ | 2~4m | 2~4m |
| $2D^2/\lambda$ | 106.67m | 320m |
| Frequency | 1GHz | 3GHz |
| Wavelength | 0.3m | 0.1m |
| $\|\beta-180°\|$ | $\leq 14.6°$ | $\leq 10.9°$ |

In TABLE II, crossing point represents the distance from the transmitter to the target center. The value of $2D^2/\lambda$ is the criteria for determining whether a target located in the near field or far field. The angle $|\beta-180°|$ indicates the bistatic angle of the moved target from 180°.

The higher frequency is deliberately selected to generate FS signals with rich information. Because spherical waves degenerate into plane waves during long-distance propagation, increasing the incident wave frequency will compensate for the lost geometric information.

This paper balances simulation accuracy and computational efficiency by setting strategies of different resolution grids. In the simulation stage, we use a high-density grid (with a resolution of $\lambda/16$) within classic FKDM to generate highly accurate reference FS signals, which serve as ground truth. During the inverse learning process, the SWRF parameters are defined on a coarser grid ($\lambda/8$), significantly reducing computational complexity while still preserving important shadow features. The multi-resolution design provides rigorous validation of the resolution generalization ability of our DPI-SPR framework. By learning the coarse-resolution SWRF parameters to reproduce the fine-resolution reference signal, we demonstrate that our model is able to learn a robust continuous physical representation rather than overfitting to specific discretized grids.

1) SWRF Parameter Initialization

A well-chosen initialization for the SWRF parameters is crucial for ensuring stable and efficient convergence. Directly guided by the foundational principle of our framework, we adopt a physical prior that incorporates the concept of soft boundary rather than a random initial state.

We initialize SWRF with a radially symmetric soft circular distribution on $\Omega_0$, which completely encloses the shadow of the target in the aperture plane. Mathematically, this soft circular prior distribution is generated using an exponential kernel function. Let the initial SWRF value at each secondary wave center $Q$ on $\Omega_0$ be defined as:

$$\boldsymbol{\rho}^{(0)} = T - \exp(\frac{d_m}{R}) \quad (21)$$

Where $d_m$ denotes the distance from each point $Q$ to the center of the plane $x'oz'$. The hyperparameter $T=10$ controls the peak activation level inside the circle, while $R$ defines the effective soft radius and controls the sharpness of the decay at the boundary. The computational domain $\Omega_0$ is a circular region with a radius of 3.2 m. The initialization strategy in Equation (21) provides a physically motivated prior assuming single-target scenarios near the aperture center, while maintaining geometric neutrality to preserve optimization freedom for data-driven shape convergence.

As illustrated in Fig. 5, this strategy consistently generates a radially symmetric distribution of the $\sigma(\rho)$ that is high at the center and decays smoothly towards the periphery.

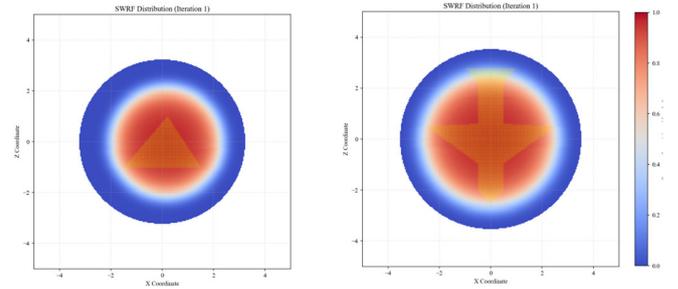

(a) Initialized small Radius  (b) Initialized big Radius

Fig. 5. The activation coefficient $\sigma(\rho)$ of initial distribution

we set $R=0.9$ for geometrically simple targets and $R=1.2$ for larger, more complex targets. The initial center of the circle does not have to be strictly limited to the origin.

2) Performance Evaluation Metrics

In this framework, the inversion aims to reconstruct a binary shadow profile representation. To objectively assess DPI-SPR performance, we evaluate two complementary dimensions: global reconstruction accuracy using the Intersection over Union (IoU) metric for pixel-level structural integrity assessment, and local geometric accuracy using the 95th percentile Hausdorff Distance (95% HD) for boundary-based profile precision evaluation.

a) 95% Hausdorff Distance

The 95% HD assesses local boundary accuracy by measuring the maximum error between two boundaries after ignoring the most extreme 5% outliers[32]. Based on HD, the 95%HD is shown by

$$HD_{95}(B_{rec}, B_T) = \max\{d_{95}(B_{rec}, B_T), d_{95}(B_T, B_{rec})\} \quad (22)$$

The 95% HD values are comparable for several grid sizes, indicating a high level of accuracy in the reconstruction of shadow boundaries.

b) Intersection over Union (IoU)

The IoU metric evaluates the overall structural integrity and spatial correspondence. It is mainly used to quantify the overlap between the reconstructed shadow profile $S_{rec}$ and the ground truth shadow profile $S_T$, and is defined as

$$\text{IoU} = \frac{|S_{rec} \cap S_T|}{|S_{rec} \cup S_T|} \quad (23)$$

IoU values approaching one indicate nearly perfect geometric correspondence for the target shadow profile reconstruction.

### B. Generality Evaluation of DPI-SPR

The generality and fundamental accuracy of the DPI-SPR framework under ideal conditions are validated in this section. We performed on four canonical targets with convex polygons, including triangle, square, trapezoid, and hexagon in TZ and FF scenarios. Numerical simulation experiments on different targets and FSR scenarios will demonstrate the general applicability of the DPI-SPR framework.

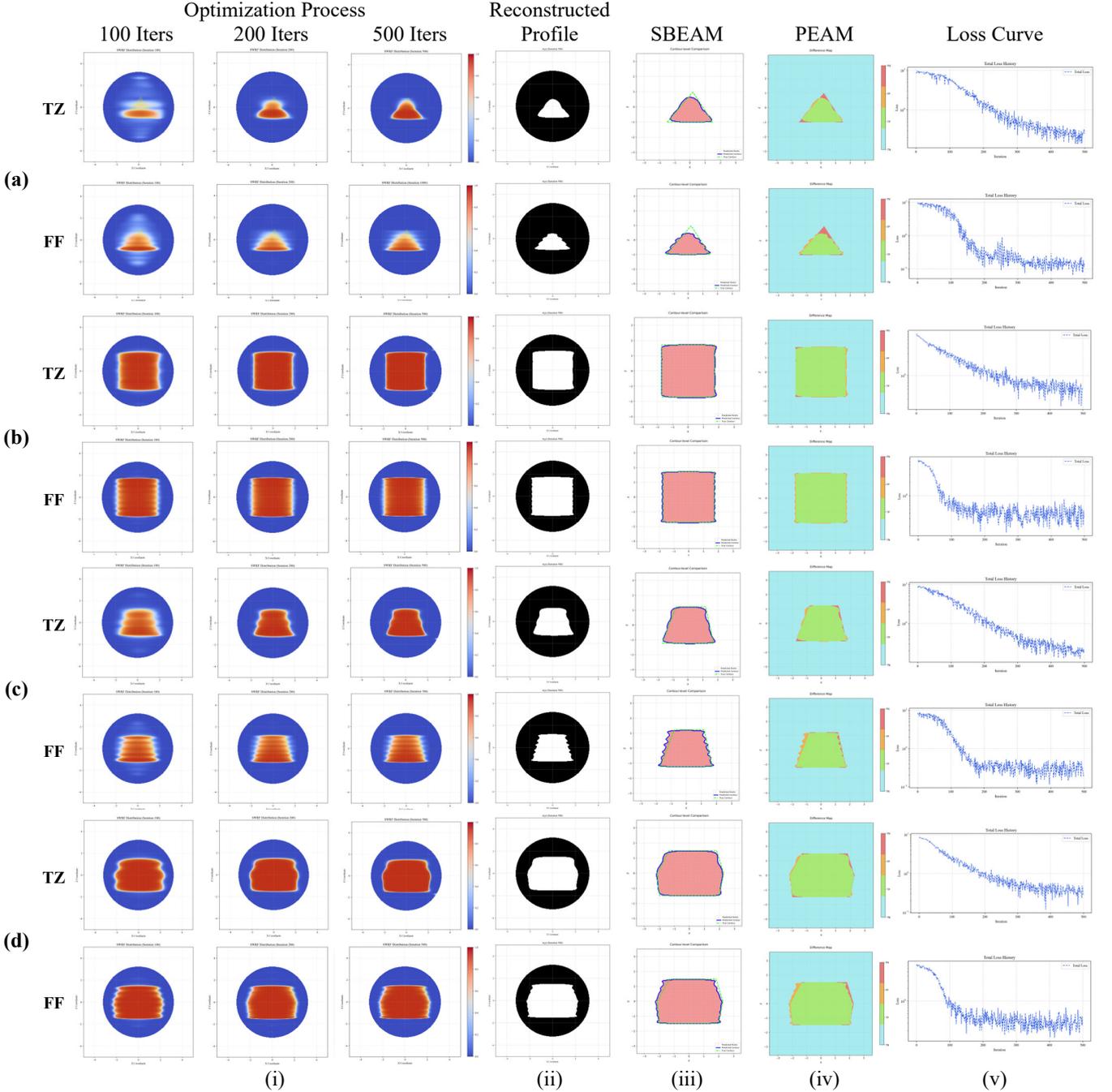

**Fig. 6.** Visualization of SWRF parameter inversion for four convex objects. (i) Optimization process of 100~500 iterations, (ii) Final reconstructed profiles, (iii) Shadow Boundary Error Analysis Map (SBEAM), (iv) Pixel-wise Error Analysis Map (PEAM),

(v) Loss curves. Rows (a), (b), (c), (d) visualized the results of triangle, square, trapezoid, and hexagon under TZ and FF scenarios, respectively.

The hyperparameter settings for DPI-SPR optimization under the AD solver are detailed in TABLE II.

TABLE II HYPERPARAMETERS FOR SIMPLE TARGETS

|    | $\delta$ | $\lambda_{Ent}$ | $\lambda_1$ | $\lambda_2$ | $\lambda_3$ |
|----|----------|-----------------|-------------|-------------|-------------|
| TZ | 40       | 0               | 0           | 0           | 0           |
| FF | 1e5      | 0               | 0           | 200         | 0           |

The AD solver adopted the Adam optimizer to learn SWRF parameters. The initial learning rate is set to be 0.05, and decays to 50% of the current learning rate every 300 iterations. The maximum number of iterations is set to 500. The distribution of activation probability on computational region is visualized and analyzed throughout the iterative learning process and upon convergence.

Fig. 6 provides a comprehensive visual narrative of the optimization process. The activation probabilities $\sigma(\rho)$ based on SWRF evolve from an initial circular prior $\rho^{(0)}$ to distributions consistent with the target shadow profiles in column(i). The corresponding loss curves exhibit rapid and stable convergence in the column(v). Among them, the loss in the FF converges faster than that in TZ, because FF training mainly selected L2 loss.

The final reconstructed results show binary shadow profiles $\Omega_{Tar}$ in column (ii). The PEAM and SBEAM analysis analyses were used to quantitatively evaluate the performance of our DPI-SPR method. The 95%HD and IoU metrics of corresponding evaluation results are illustrated in TABLE III. It is evident from columns (ii) to (iv) that the experimental results in all test cases exhibit a high degree of geometric consistency with the true binary profiles. The union of all true positive (TP) and false negative (FN) pixels in column (iv) represents the true shadow profile and the green line.

TABLE III PERFORMANCE EVALUATION FOR FOUR TARGETS

|       |    | Triangle | Square | Trapezoid | Hexagon |
|-------|----|----------|--------|-----------|---------|
| IoU   | TZ | 0.8786   | 0.9631 | 0.9359    | 0.9638  |
|       | FF | 0.8197   | 0.9578 | 0.9132    | 0.9408  |
| 95%HD | TZ | 0.1891   | 0.0781 | 0.1431    | 0.1352  |
|       | FF | 0.2755   | 0.0853 | 0.2022    | 0.1856  |

TABLE III shows that the IoU scores consistently exceed 0.81, with the majority results surpassing 0.91, indicating a high level of accuracy in global reconstruction. The main reason for the lower reconstruction results of triangular shadow is the imprecise SWRF parameter inversion near the vertices. The 95% HD values with the range of 0.26~1λ confirmed that the reconstructed shadow profiles have a high level of imaging accuracy.

A significant finding is that TZ consistently outperforms FF on all targets. This is because TZ observations utilize the spherical wavefront geometry, which contains comprehensive phase and amplitude constraints, thereby providing stronger inverse problem conditions. The performance difference highlights the theoretic advantage of near-field measurements and the robust adaptability of our DPI-SPR in various observation scenarios, demonstrating the generalization capability of the underlying framework.

*C. Robustness Evaluation of DPI-SPR*

After demonstrating the applicability of our framework, we turn to examining its robustness under realistic operating conditions. In real FSR systems, signal corruption is a common issue due to additive noise and secondary scattering artifacts, which can manifest as outlier-like perturbations. Therefore, it is crucial for FSR imaging techniques to maintain robust performance in the face of these perturbations for practical deployment. This section evaluates the effectiveness of our DPI-SPR in reconstructing shadow profiles using SWRF parameters which is directly inverted from noisy signals in both TZ and FF scenarios, specifically for non-convex targets with more complex geometries.

All reconstructed profiles in this section were achieved by using a set of fixed hyperparameters on three different targets, as listed in Table X. The maximum number of iterations of the optimization process is 1000.

TABLE IV HYPERPARAMETERS FOR COMPLEX TARGETS

|       |       | $\delta$ | $\lambda_{Ent}$ | $\lambda_1$ | $\lambda_2$ | $\lambda_3$ |
|-------|-------|----------|-----------------|-------------|-------------|-------------|
| Clean | TZ    | 30       | 0.5             | 0.1         | 0           | 0.1         |
|       | FF    | 30       | 0.5             | 0.1         | 200         | 0.1         |
| Noise | TZ/FF | 1e5      | 1.0             | 0.1         | 270         | 0.1         |

The reference signal $y[k]$ in a noisy environment is expressed as

$$y[k] = s[k] + n_G[k] + n_I[k] \qquad (24)$$

Where $s[k]$ is the noiseless FS signal simulated by FKDM. The term $n_G[k] \sim N(0, \sigma_G^2)$ represents additive white gaussian noise (AWGN) and the term $n_I[k]$ indicates that a high-amplitude pulse noise occurs with a probability of $p_i = 5\%$. The power of total noise is

$$P_N = \sigma^2 + \alpha^2 P_s p_i \qquad (25)$$

The impulse noise magnitude is scaled by $\alpha = 1.5$ relative to the root-mean-square amplitude of the $s[k]$, where $P_s$ represents the power of FS signal. The Optimization process and reconstructed shadow profiles derived from converged SWRF parameters are presented in Fig. 6 and Fig. 7. TABLE V presents quantitative assessment of our DPI-SPR method on noisy FS signals.

The qualitative results for Fig. 7 intuitively confirmed the flexibility and robustness of the DPI-SPR framework. Column(i) visualized the convergence of the optimization process from an initial circle prior (Fig. 5(b)) to correct complex shapes for all objects. Even under severe mixed noise conditions, the final reconstructed binary profiles still maintained coherence and topological correctness. The corresponding loss curves also show stable convergence despite oscillations caused by noise.

|   | Optimization Process | | | | Reconstructed Profile | SBEAM | PEAM | Loss Curve |
|---|---|---|---|---|---|---|---|---|
|   | 100 Iters | 200 Iters | 500 Iters | 1000 Iters | | | | |

(1) (a), (b), (c), (d)

(2) (a), (b), (c), (d)

(3) (a), (b), (c), (d)

(i) (ii) (iii) (iv) (v)

**Fig. 7**. The robustness analysis of the DPI-SPR framework against various noise conditions for non-convex targets under TZ observation. For each target, the inversion process is evaluated across four distinct noise environments in block (1)(2)(3). Within each block, the four rows correspond to the following conditions, from top to bottom: (a) the ideal noiseless case, (b) the SNR of FS signal with AWGN is 15dB, (3) the SNR of FS signal with AWGN is 8dB, and (d) the SNR of FS signal with AWGN and impulse noise is 8dB. Columns(i)~(v) are the same as those in Fig. 5.

TABLE V RESULTS FOR NON-CONVEX TARGETS (TZ)

|  | SNR | Tar-1 | Tar-2 | Tar-3 |
|---|---|---|---|---|
| IoU | Infinite | 0.8804 | 0.9386 | 0.9247 |
|  | 15dB | 0.8385 | 0.8936 | 0.8741 |
|  | 8dB | 0.8301 | 0.8664 | 0.8457 |
|  | 8dB-Mixed | 0.8344 | 0.8485 | 0.8351 |
| 95%HD | Infinite | 0.2336 | 0.1243 | 0.0922 |
|  | 15dB | 0.3250 | 0.2419 | 0.1668 |
|  | 8dB | 0.2820 | 0.2922 | 0.2062 |
|  | 8dB-Mixed | 0.3234 | 0.2668 | 0.2155 |

The quantitative analysis presented in Table V demonstrates the robustness of the framework by evaluating IoU and 95%HD values. In noise-free conditions, the framework performs exceptionally well, achieving an IoU value of 0.9386 for Tar-2 and exceeding 0.88 for the remaining objects. This demonstrated its ability to handle non-convex geometry of complex target. As the SNR decreases, the IoU decreases in a predictable monotonic manner. Interestingly, the impact of mixed noise scenarios (8dB mixed noise vs. 8dB Gaussian noise) is minimal, with the IoU for Tar-2 only decreasing from 0.8664 to 0.8485, and the values for the remaining two targets experiencing only minor changes. All experimental results demonstrated that our method can remove outliers and maintain the overall structure integrity of the shadow profile.

Notably, the 95% HD metric exhibited counterintuitive behavior in relation to the variation of SNR. For Tar-1, the boundary accuracy is better at 8dB (0.2820) compared to 15dB (0.3250) significantly, while for Tar-2, the performance is better under mixed noise (0.2668) compared to purely ADWN (0.2922). This apparent irregularity can be attributed to the potential for noise to act as a stochastic regularization when optimizing over a complex loss graph. This can lead to the SWRF parameters converging to a more optimal solution. For instance, moderate noise levels can trap the optimization in suboptimal local minima containing local boundary artifacts. However, stronger or mixed noise perturbations can provide random "shocks" that help escape these traps and enable convergence to a more optimal minimum with a more uniform error distribution.

A similar validation further confirmed the robustness of the DPI-SPR framework under information-sparse conditions in FF observation as shown in Fig. 8. Despite the presence of severe noise perturbations, main experimental results also demonstrated that even from an initial circular prior (Fig. 5(b)), the optimization processes converged to the correct shadow profile under noisy signals. The quantitative evaluation results with IoU and 95%HD are listed in TABLE III.

TABLE IV RESULTS FOR NON-CONVEX TARGETS (FF)

|  | SNR | Tar-1 | Tar-2 | Tar-3 |
|---|---|---|---|---|
| IoU | Infinite | 0.8247 | 0.8524 | 0.8353 |
|  | 15dB | 0.7971 | 0.8310 | 0.8506 |
|  | 8dB | 0.8074 | 0.8740 | 0.8219 |
|  | 8dB-Mixed | 0.6764 | 0.9002 | 0.8514 |
| 95%HD | Infinite | 0.2965 | 0.2516 | 0.1767 |
|  | 15dB | 0.2854 | 0.1853 | 0.1710 |
|  | 8dB | 0.3081 | 0.1644 | 0.2039 |
|  | 8dB-Mixed | 0.3972 | 0.1480 | 0.1690 |

For the targets Tar-2 and Tar-3, noise acts as a stochastic regularization, resulting in unexpected improvements in performance. Mixed noise conditions lead to a 3% increase in IoU values compared to ADWN, as shown in TABLE VI. This improvement is also evident in Fig. 8, where the reconstructed profiles exhibit better overlap fidelity in both SPEAM and PEAM analyses, despite oscillatory loss trajectories in mixed noise. This phenomenon is also observed in the TZ scenario. We hypothesize that the impulsive perturbation adds a stochastic regularization, allowing the optimizer to escape suboptimal solutions and explore a wider range of the global solution space.

However, the performance of the Tar-1 target decreases significantly under mixed noise, with a 15% decrease in IoU value and an increase in 95%HD by one λ. This is likely due to the limited information available in the FF scattering signal from smaller lateral targets, which is further masked by noise interference. These experimental results highlight a fundamental threshold effect, where noise is possible to optimize information utilization, but a minimum level of information content is necessary for accurate shadow profile inversion. This is supported by the consistently superior performance of the Tar-1 target in the TZ scenario (Fig. 7(1)), where the rich near-field information satisfies the minimum physical information required for reconstruction. The minimum loss value after convergence is mainly affected by the noise energy, and the final convergence of the loss curve indicates that our framework has good robustness. With the exception of Tar-1, the visualized reconstructions closely resemble the true shadow profile, indicating that our method converges to a physically plausible solution rather than overfitting to the noise.

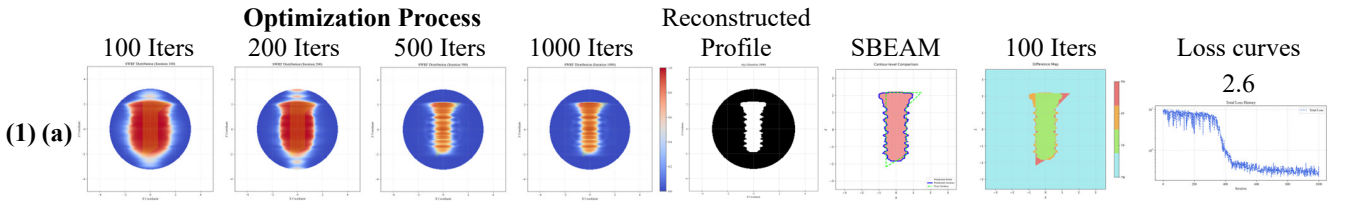

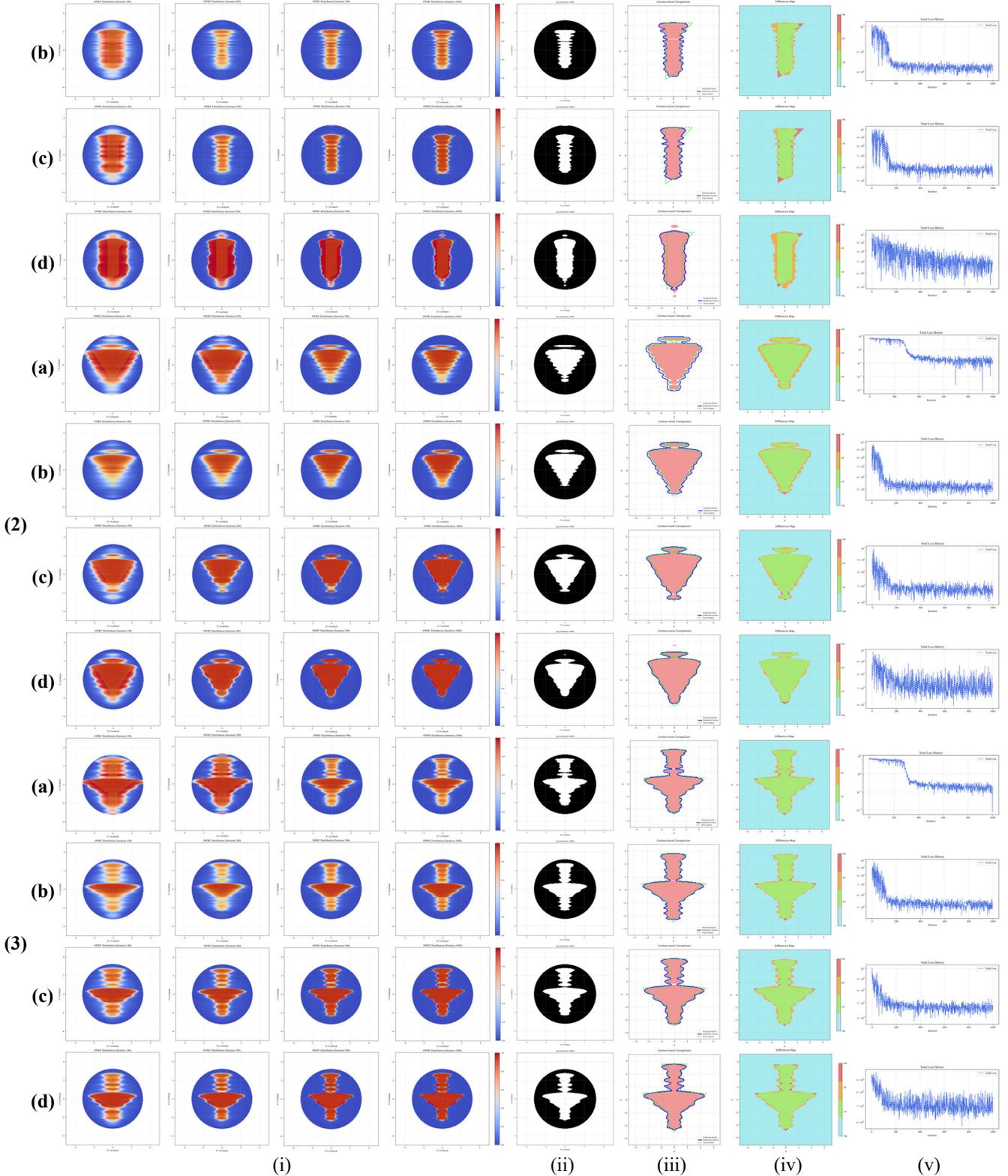

**Fig. 8**. The robustness analysis of the DPI-SPR framework against various noise conditions for non-convex targets under FF observation. For each target, the inversion process is evaluated in block (1)(2)(3), which are the same as Fig. 7.

The successful inversion of target SWRF parameters without relying on independent hyperparameter settings fully demonstrated the robustness and generalizability of this framework.

To verify physical faithful of our DFSM, we conducted signal domain fidelity analysis as presented in Fig. 9. This assessment aimd to verify whether the FS signals generated by the reconstructed binary profiles in Fig. 7 and Fig. 8 are consistent with the true values of the observed signals, thereby validating the physical consistency of the inversion results.

Fig. 9(a) and (d) show the variation characteristics of the FS signal for Tar-3 under ideal TZ and FF scenarios, respectively. FS signals of the binary profiles predicted and reconstructed using the FKDM (dashed red line) showed excellent data consistency with the ideal ground truth FS signal (solid blue line), confirming that the shadow profile reconstructed by our method is physically reliable in accordance with the FSR diffraction principle.

In addition, this paper evaluated the performance of the DPI-SPR framework under severe noise corruption conditions. For example, under 8dB mixed noise interference, the referenced FS signals (solid blue line) shown in Fig. 9(b) and (e) exhibited severely anomalies, but the framework successfully extracted the underlying physical information. The predicted signal (dashed red line) maintains a smooth characteristic consistent with the blue curve.

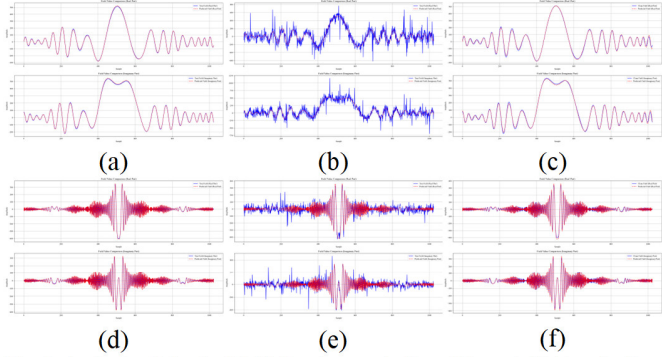

**Fig. 9.** Analysis of physical Self-Consistency in Signal Domain Taking the Tar-3 target as an example, The top row (a-c) corresponds to the TZ scenario, and the bottom row (d-f) corresponds to the FF scenario.

The key comparisons in Fig. 9(c) and (f) demonstrate that the final predicted FS signals are highly consistent with the ground-truth signal. While slight deviations occur near peaks of FS signals, the overall consistency remained high, demonstrating the excellent noise resilience of our method.

Analysis ultimately validates the noise resilience of our DPI-SPR framework from signal domain, specifically its ability to recover the target shadow profile, including the underlying diffraction features, from severely degraded measured FS signals, making it suitable for practical FSR applications.

### D. Ablation Studies

This section presents a series of ablation experiments aimed at quantitatively validating the contributions of each key component proposed in the inverse parameter learning process of the DPI-SPR framework. The experiments primarily focus on investigating the necessity of the logarithmic loss function, entropy regularization term, and geometric regularization prior through simulation. All experiments are conducted in the FF scenario, where effective information is limited. The SNR of the mixed noise signal used for SWRF parameter learning is 8dB for Tar-2 and Tar-3. The reconstructed results of our method are used as a baseline and compared with the results of multiple ablation experiments.

TABLE VI ANALYSIS OF ABLATION STUDY RESULTS

|  |  | (b) | (c) | (d) | (e) | (f) | (g) | (h) |
|---|---|---|---|---|---|---|---|---|
| 95% HD | Tar-2 | 0.157 | 0.996 | 0.565 | 0.272 | 0.190 | 0.290 | 0.269 |
|  | Tar-3 | 0.188 | 2.197 | 2.171 | 0.279 | 0.232 | 0.237 | 0.237 |
| IoU | Tar-2 | 0.892 | 0.611 | 0.656 | 0.837 | 0.871 | 0.838 | 0.828 |
|  | Tar-3 | 0.847 | 0.457 | 0.483 | 0.741 | 0.784 | 0.803 | 0.831 |

Fig. 10(a) in the first and third rows showed the initial distribution of SWRF parameters, with the yellow geometry representing the ground-truth target shadow profile. TABLE VII and Fig. 10(b) show the reconstructed shadow profiles of the DPI-SPR framework, demonstrating extremely high reconstruction accuracy, both visually and in terms of the explicit 95% HD and IoU values.

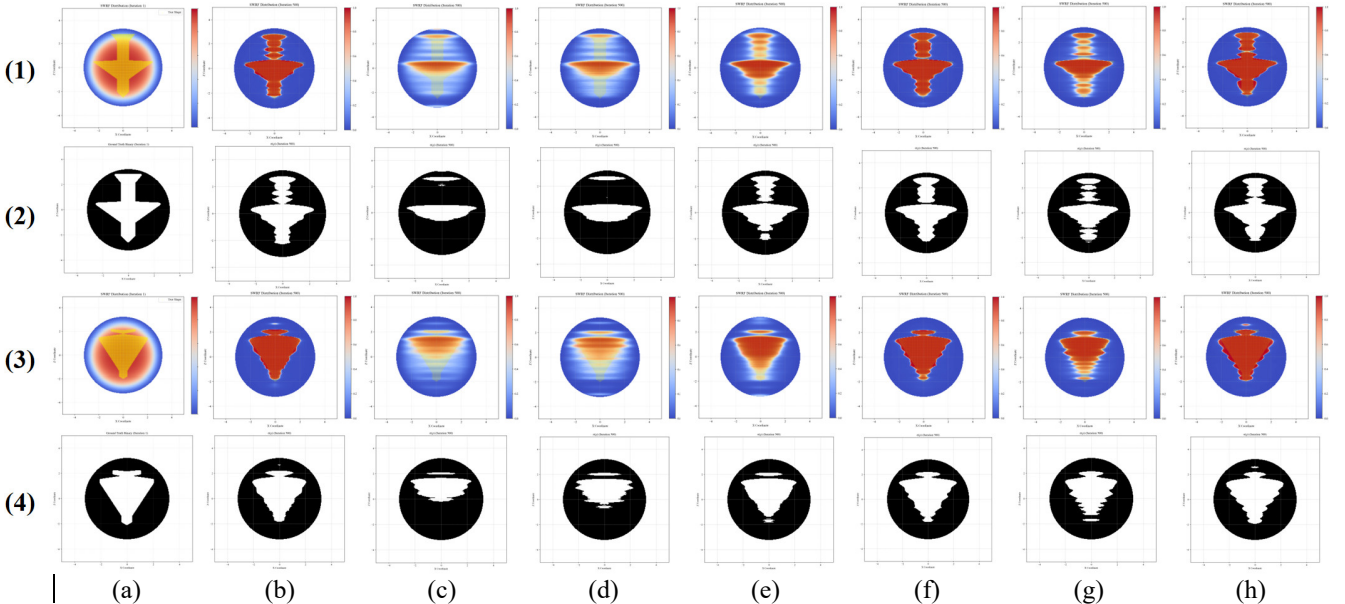

**Fig. 10**. Ablation study results demonstrating the indispensable role of each component in the DPI-SPR framework. For each target, the top row visualizes the $\sigma(\rho)$ of converged SWRF field, while the bottom row shows the final binary profile, both under 500 iterations. Each column represents a different model configuration. (a) Ground Truth (GT). (b) Full DPI-SPR Model (Baseline). (c) Without Logarithmic Loss (using a standard L2 Loss). (d) Without any Regularization

(Logarithmic Loss only). (e) Without Geometric Regularization (Logarithmic Loss and Entropy Regularization. (f) Without Compactness Regularization. (g) Without Connectivity Regularization. (h) Without Smoothness Regularization.

In contrast, other ablation models visualize the activation probabilities of the inverted SWRF parameters as (1), (3) and the reconstructed shadow outlines as (2), (4) in Fig. 10(c)~(h). It can be clearly seen that the target geometry accuracy has degraded, highlighting the necessity of designing each component in the DPI-SPR framework.

In Fig. 10(c), replacing the logarithmic loss function with the standard L2 loss function results in catastrophic failure. This demonstrates that the optimization becomes unstable in the presence of impulsive noise, causing the shadow outlines to become completely distorted and unrecognizable. This confirms the importance of the logarithmic loss function in handling outliers and strong noise, promoting stable and rapid convergence of the model. The results in Fig. 10(d), (e) clearly demonstrate the critical role of the regularization term. When relying solely on the logarithmic loss function without any regularization, the result in Fig. 10 (d) leads to severe distortion, highlighting the limitations of relying solely on the physical DFSM to constrain this ill-posed problem.

However, adding entropy regularization in Fig. 10(e) significantly improves the inversion results, demonstrating that maintaining the liveness of the gradient effectively prevents the optimizer from prematurely falling into incorrect and unreasonable states, thereby stabilizing the optimization process. Although the addition of entropy regularization makes the inversion results closer to the true projective geometry, most activation probabilities $\sigma(\rho)$ of the inverted SWRF parameters in Fig. 10(2)(e) and (4)(e) are distributed around 0.5, which lacks true physical rationality. In addition, the reconstructed binary shadow contours appear to be disconnected components rather than clear and coherent boundaries, which indicates that it is necessary to use prior geometric features as regularization.

In Fig. 10(f)~(h), removing each regularization component reveals the respective effects of the three regularizations. Removing the connectivity prior causes the contour to become fragmented compared to the baseline result Fig. 10(b), highlighting its critical role for target shadow profile. Similarly, removing the compactness or smoothness components will cause slight geometric distortions in the inverted contours, thereby reducing the inversion accuracy.

Our detailed analysis of the ablation experiments reveals the different contributions of each component in the framework. The logarithmic loss provides noise stability, the binary entropy regularization enables it to escape from local optimality, and the connectivity prior enforces the single coherent target constraint, which constitutes an important basis for successful inversion. On this basis, the smoothness and compactness priors serve as refinement mechanisms to promote the sharpening of profile boundaries, thereby improving the inversion accuracy. In summary, the DPI-SPR framework ensures that reconstructed profiles are physically reasonable and geometrically accurate.

## V. Conclusion

In this work, we presented DPI-SPR, a differentiable physical inversion paradigm for robust shadow-profile reconstruction in FSR. The method reformulates traditional analytical and pixel-based retrieval as an optimization over learnable SWRF parameters. By modeling the target shadow as a continuous, learnable physical field, DPI-SPR yields a fully differentiable framework that accommodates the complexity of real-world scattering.

Through extensive numerical simulations, we demonstrated that DPI-SPR accurately reconstructs shadow profiles of complex targets, even under low SNR with mixed noise. Notably, the framework remained highly robust at SNR as low as 8dB and performed exceptionally well in the TZ.

The robustness and effectiveness of proposed DPI-SPR framework are attributed to its integration of four key elements: (1) a differentiable, unconstrained SWRF parameterization for modeling complex target geometries; (2) a physics-informed forward model that preserves wave propagation characteristics; (3) a robust logarithmic loss function to mitigate the impact of outliers; and (4) geometric regularization priors that guide the optimization towards physically coherent solutions.

In summary, DPI-SPR provided an accurate, robust, and automated solution for shadow profile reconstruction in FSR. It sets a strong baseline for future FSR systems and offers a general approach to inverse scattering. Future work will include real-world validation, extensions to multi-target scenarios, and exploration of full 3D reconstruction.